\documentclass[11pt,twoside]{article}

\usepackage{asp2014}

\aspSuppressVolSlug
\resetcounters

\begin{document}

\title{A hybrid architecture for astronomical computing}

\author{Changhua Li$^1$, Chenzhou Cui$^1$, Boliang He$^1$, Dongwei Fan$^1$,Linying Mi$^1$,Shanshan Li$^1$, Sisi Yang$^1$, Yunfei Xu$^1$, Jun Han$^1$,Junyi Chen$^2$, Hailong Zhang$^3$, Ce Yu$^4$, Jian Xiao$^4$, Chuanjun Wang$^2$, Zihuang Cao$^1$, Yufeng Fan$^2$, Liang Liu$^6$, Xiao Chen$^7$, Wenming Song$^7$, Kangyu Du$^8$
\affil{$^1$National Astronomical Observatories, Chinese Academy of Sciences (CAS), 20A Datun Road, Beijing 100012, China; \email{lich@nao.cas.cn}}
\affil{$^2$Yunnan Astronomical Observatory, CAS, P.0.Box110, Kunming 650011, China; \email{wcj@ynao.ac.cn}}
\affil{$^3$Xinjiang Astronomical Observatory, CAS, 150 Science 1-Street, Urumqi, Xinjiang 830011, China; \email{zhanghailong@xao.ac.cn}}
\affil{$^4$Tianjin University, 92 Weijin Road, Tianjin 300072, China; \email{xiaojian@tju.edu.cn}}
\affil{$^5$Purple Mountain Observatory, 2 West Beijing Road, Nanjing 210008, China; \email{liangliu@pmo.ac.cn}}
\affil{$^6$Shanghai Astronomical Observatory, 80 Nandan Road, Shanghai 200030, China ; \email{cx@shao.ac.cn}}
\affil{$^7$Beijing University Of Technology, Beijing 100124, China ; \email{songwenming@nao.cas.cn}}
\affil{$^8$Central China Normal University, 152 Luoyu Road, Wuhan 430079, China;\email{dukangyu@nao.cas.cn}}
}

\paperauthor{Changhua Li}{lich@nao.cas.cn}{}{National Astronomical Observatories, Chinese Academy of Sciences (CAS)}{Center of Information and Computing}{Beijing}{Beijing}{100012}{China}
\paperauthor{Chenzhou Cui}{ccz@bao.ac.cn}{}{National Astronomical Observatories, Chinese Academy of Sciences (CAS)}{Center of Information and Computing}{Beijing}{Beijing}{100012}{China}

\begin{abstract}
	With many large science equipment constructing and putting into use, astronomy has stepped into the big data era.  The new method and infrastructure of big data processing has become a new requirement of many astronomers. Cloud computing, Map/Reduce, Hadoop, Spark, etc. many new technology has sprung up in recent years. Comparing to the high performance computing(HPC), Data is the center of these new technology. So, a new  computing architecture infrastructure is necessary, which can be shared by both HPC and big data processing.  Based on Astronomy Cloud  project of Chinese Virtual Observatory (China-VO),  we have made much efforts to optimize the designation of the hybrid computing platform. which include the hardware architecture, cluster management, Job and Resource scheduling.
\end{abstract}

\section{Introduction}
With the large-scale astronomical data generation and astronomical numerical simulation scale expansion, all of these are highly dependent on high-performance computing power. Therefore, building an efficient computing infrastructure has become an indispensable part of astronomical research. The improvement of computing performance not only depends on hardware processing ability, scheduling and management of cluster play a important role.

\section{GPU computing}
Recently 10 years, GPU have become widely used nowadays to accelerate a broad range of applications, including computational physics and astrophysics, image/video processing, engineering simulations, quantum chemistry. Researchers and developers have enthusiastically adopted CUDA and GPU computing for a diverse range of applications, published hundreds of technical Papers and teached CUDA programming at more than 300 universities\citep{gpueraL}. GPU computing has advanced from a specialists' branch to mainstream in supercomputing. NVIDIA GPUs is the Representative example. In 2009, "Laohu", a major GPU cluster was designed and installed at National Astronomical Observatoris, Chinese Academy of Sciences(NAOC). On this system, we reach sustained speeds for our real astrophysical N-body application of 50 Teraflops\citep{Berczik2013Up}. This is a very good performance compare to the pure CPU program.
\articlefigure{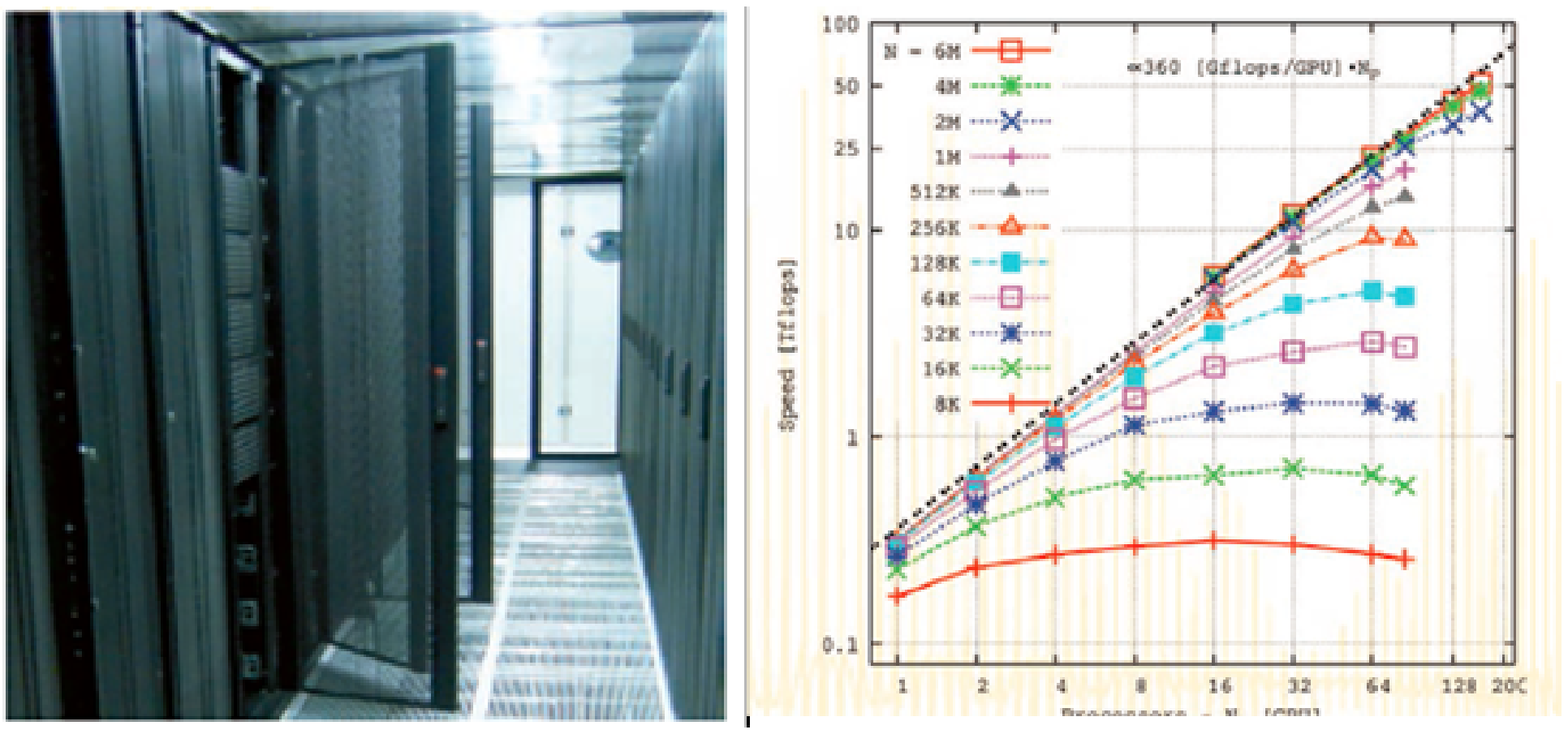}{f1}{Laohu cluster and the bentch mark result}

\section{Intel Knights Landing}
Knights Landing(KNL) is the second-generation Intel Xeon Phi product family, which targets high-performance computing and other highly parallel workloads. Compare to the GPU or Knights Corner, the GPU or Knights Corner only is the coprocessor, which must depends on the CPU of host server, but Knights Landing is a really processor with 72 cores and integrates high bandwidth memory known as MCDRAM which greatly enhances performance. Especially, it supports the same parallel programming models, the same tools, and the same binaries that run today on other Intel processors, which is a good news for all astronomers.

\section{Cloud computing}
	Cloud computing has many definitions, here mainly refers to HADOOP\citep{hadoop}, SPARK\citep{Spark} and other large data processing framework, which will greatly accelerate the astronomical data processing and analysis. In 2015,
	under funding support from NDRC (National Development and Reform commission) and CAS (Chinese Academy of Sciences), astronomy cloud computing platform\citep{P1-2_Li_adassxxiv} was released for all astronomy researcher. Astronomy cloud computing platform includes rich and geographical distribution's computing, storage resource and provide online custom configuration service. Based on this platform, virtual computing cluster can be built, each computing node is a virtual machine which support the spark, map/reduce parallel computing model and HDFS file system.

\section{A new hybrid computing platform}
	
	In order to meet the requirement of the diversity of astronomical computing and increase the resource utilization of computing infrastructure, a new hybrid architecture computing platform was designed and implemented in China-VO, which integrated cloud computing, GPU and KNL cluster. Figure 2 shows the architecture.
\articlefigure{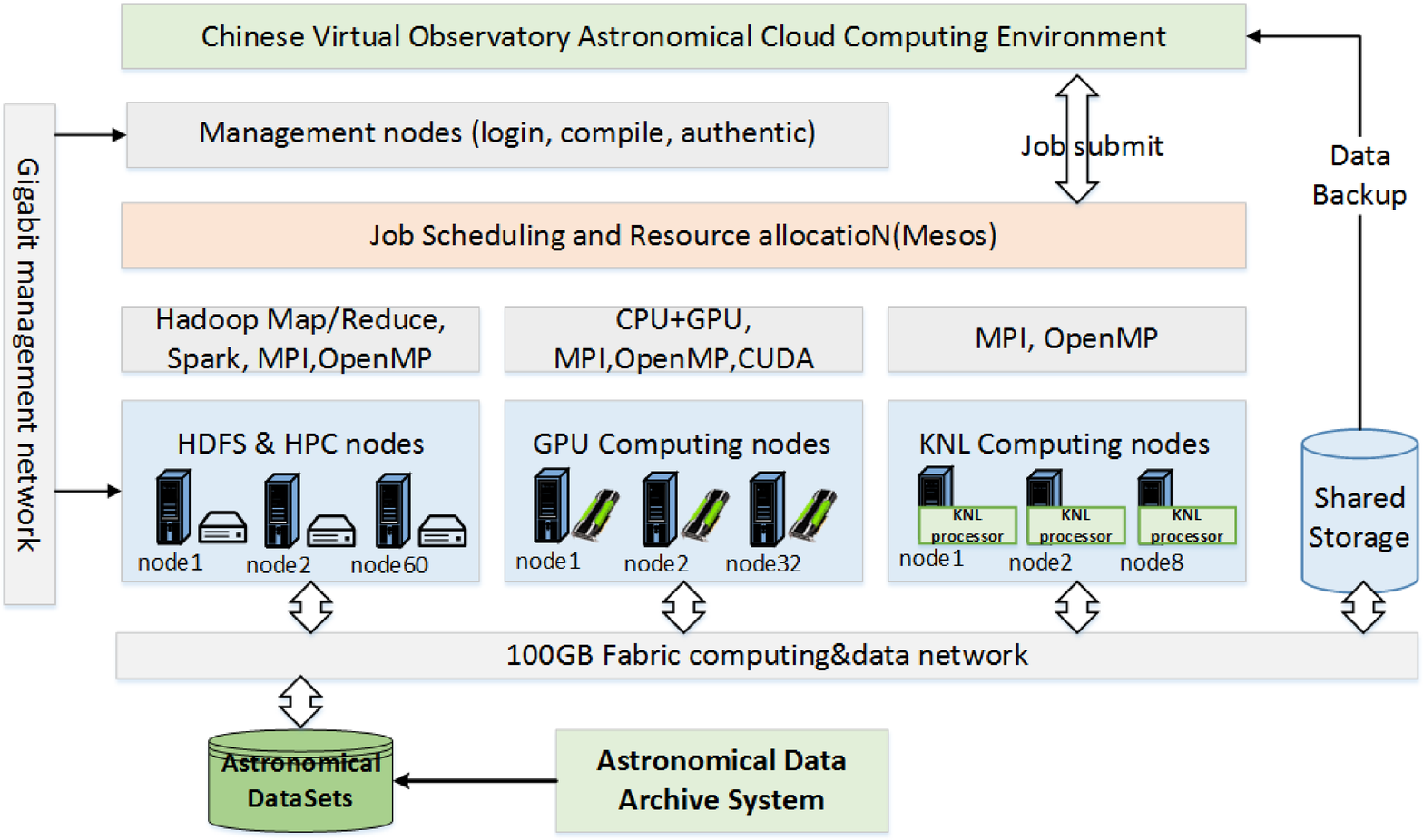}{f1}{the architecture of hybrid computing}

	In this architecture, Computing resource was divided two layer. The top layer is the cloud computing, which receive user's job and resource requirement, responsible for user management and task allocation.  User can choose computing resource type. The second layer is the classic HPC layer, includes computing server cluster, GPU cluster and Knights landing cluster. According to the requirement of users. Resource scheduling component can send computing task to special resource cluster, monitoring task running status and return the result, so, resource scheduling and management is the key part. 
 The astronomical dataset and shared storage system can be shared by all type computing nodes.

\section{Conclusion}
	The integration of cloud computing and high-performance	computing poses a challenge to cluster management such as job management and scheduling and resource allocation,which greatly increases the complexity of these software.
	But meanwhile, this integration increase the flexibility of cluster operation, and can support a variety of astronomical computing tasks, improve the efficiency of the cluster.

\acknowledgements
	This paper is funded by National Natural Science Foundation	of China (11503051, U1231108, U1531246, U1531115), Ministry of Science and Technology of China (2012FY120500), Chinese Academy of Sciences (XXH12503-05-05).
	Data resources are supported by Chinese Astronomical Data Center.

\bibliography{P37}  

\end{document}